\documentclass[11pt,a4paper]{article}
\usepackage[latin1]{inputenc}
\usepackage[english]{babel}
\usepackage{graphicx}
\usepackage{mathtext}
\usepackage{pstricks}
\usepackage{color}
\usepackage{epsfig,cite}
\usepackage{amsmath}
\usepackage{amssymb}
\usepackage{amsfonts}
\usepackage{latexsym}
\usepackage{wasysym}
\usepackage{bm}
\catcode`@=11 

\newcommand{\beq}{\begin{equation}}
\newcommand{\eeq}{\end{equation}}
\newcommand{\bea}{\begin{eqnarray}}
\newcommand{\eea}{\end{eqnarray}}
\newcommand{\bdm}{\begin{displaymath}}
\newcommand{\edm}{\end{displaymath}}
\def\as{\alpha_s}

\begin{document}
\pagestyle{empty}

\begin{flushright}
MAN/HEP/2011/09 \\
\end{flushright}

\begin{center}
\vspace*{2.5cm}
{\Large \bf Accurate QCD predictions for new variables to study dilepton transverse momentum$^*$}
 \\
\vspace{0.3cm}
Simone Marzani$^a$, Andrea Banfi$^{b}$, Mrinal Dasgupta$^a$ and Lee Tomlinson$^a$
\\
\vspace{0.3cm}  {\it
{}$^a$School of Physics \& Astronomy, University of Manchester,\\
Oxford Road, Manchester, M13 9PL, England, U.K.\\ \medskip
{}$^b$Institute for Theoretical Physics, ETH Zurich, \\
        8093 Zurich, Switzerland}\\ 
\vspace*{1.5cm}

{\bf Abstract}
\end{center}

\noindent

We report on the computation of the angle-between-leptons distribution in Drell-Yan processes. More precisely we study the recently introduced variable $\phi^*$, which provides us with a more accurate probe of the low $Q_T$ domain of $Z$ boson production at hadron colliders.
Our theoretical prediction is obtained by matching a next-to--next-to leading logarithmic (NNLL) resummation to a fixed order calculation at next-to-leading order (NLO). 
We find that the result significantly differs from a pure fixed order calculation in a wide range of the observable we are interested in, clearly indicating the need for resummation.
We also perform a first comparison to the measurement by the D\O $ $ collaboration, finding good agreement between theory and experiment.
\vspace*{1cm}

\vfill
\noindent

\begin{flushleft} $^*$Presented at the XIX International Workshop on Deep-Inelastic Scattering and Related Subjects (DIS 2011), April 11-15, 2011; Newport News, Virginia, USA
\end{flushleft}
\eject

\setcounter{page}{1} \pagestyle{plain}


\section{Introduction}
The production of a lepton pair in hadron-hadron collisions is one of the most studied processes in particle phenomenology, with the original paper appearing more than forty years ago~\cite{DY}.  Since then a huge theoretical effort has gone into improving the accuracy of the predictions. For instance, QCD corrections are known to $O(\alpha_s^2)$~\cite{DYNNLO}. In particular, the transverse momentum distribution of the lepton pair, or equivalently of the gauge boson decaying into it, is of great interest. It is sensitive to multi-gluon emission from the initial state partons. This is a classical example of a multi-scale problem and the correct treatment of these effects goes beyond fixed order perturbation theory. Let us introduce the invariant mass of the leptons $M$, that will be chosen around the Z mass, and the let us call $Q_T$ the magnitude of the Z transverse momentum.
We have to consider three different regimes. When $Q_T\sim M$ we expect fixed order perturbation theory to work and programs like MCFM~\cite{mcfm} will give a good description of the process. In the region $\Lambda_{{\rm QCD}}\ll Q_T\ll M$, we can still rely on perturbation theory but large logarithms of the ratio $Q_T/M$ may spoil the convergence of the perturbative expansion and must be resummed to all orders. Finally, in the region $Q_T\sim \Lambda_{{\rm QCD}}$ we expect non-perturbative effects to play a significant role.
Therefore, it is important to compute a solid perturbative prediction, so that we can compare it to precise data coming from the experiments and be able to pin down non-perturbative contributions, related to the intrinsic transverse momentum of the initial-state quarks.  An accurate theoretical description of the transverse momentum of the Z boson is also relevant for the extraction of the W mass.

The resummation of large logarithms in the $Q_T$ spectrum has been studied by several groups for many years and it is currently known to NNLL accuracy (see Ref~\cite{FlorenceQT} and references therein). Accurate theoretical predictions have been compared to data coming from the D\O $ $ and CDF experiments at the Tevatron but no clear conclusions have been drawn to date about the relevance of non-perturbative effects. One of the limiting factor has been the experimental resolution which affects the measurement of transverse momenta. For this reason novel variables have been introduced in~\cite{WV,WVBRW} and recently measured by the D\O $ $ collaboration~\cite{D0dphi}. These variables, labelled the $a_T$ and $\phi^*$, both crucially depend on the azimuthal angle $\Delta \phi$ between the final state leptons, at low $Q_T$. The experimental resolution for $a_T$ and $\phi^*$ is significantly better than the one for $Q_T$~\cite{WVBRW}. 

The D\O $ $ collaboration compared the result of their measurement to the theoretical prediction of the program RESBOS~\cite{RESBOS}. They found overall agreement with some discrepancies in the large rapidity region. In particular, the data disfavour current non-perturbative models, such as small-$x$ broadening~\cite{BNLY} .
Therefore we need an accurate theoretical prediction for these new variables, along the line of those for $Q_T$ resummation, to be able to assess the importance of non-perturbative effects.

\section{The $\phi^*$ distribution}

In Ref.~\cite{BDMphi} we have computed a theoretical prediction for the $\phi^*$ distribution by matching a resummed NNLL calculation, which captures the dominant behaviour at small $\phi^*$ to a NLO one obtained from the program MCFM:
\begin{equation} \label{matched}
\left(\frac{ {\rm d} \sigma}{{\rm d} \phi^*}\right)_{\mathrm{matched}} = \left(\frac{{\rm d} \sigma}{ {\rm d} \phi^*}\right)_{\mathrm{resummed }} +\left(\frac{{\rm d} \sigma}{{\rm d} \phi^*}\right)_{\mathrm{NLO}}-\left(\frac{{\rm d} \sigma}{{\rm d} \phi^*}\right)_{\mathrm{expanded}}
\end{equation}
 The resummed distribution has the following form 
\bea
\label{eq:resummed}
\frac{ {\rm d} \sigma}{ {\rm d} \phi^*} \left( \phi^*,M, \cos \theta^*, y \right) &=& \frac{\pi \alpha^2}{s N_c} 
\int_0^{\infty} d b M \cos \left(bM \phi^* \right) 
e^{-R(\bar{b},M, \mu_Q,\mu_R)} \\ \nonumber &\times&\Sigma \left(x_1,x_2,\cos\theta^*, b,M,\mu_Q,\mu_R,\mu_F \right)\,,  
\eea
where
$x_{1,2} = \frac{M}{\sqrt{s}}e^{\pm y}$ and $\bar{b}= \frac{b e^{\gamma_E}}{2}$.
The above result is yet to be integrated over the dilepton invariant mass $M$, the scattering angle $\theta^*$ and rapidity of the dilepton system (or equivalently the $Z$ boson rapidity) $y$. Note the dependence on three arbitrary scales: renormalisation, factorisation and resummation, which can be varied in order to estimate the theoretical uncertainty. At the moment we set them all equal to each other and to the dilepton invariant mass $M$.

The dependence upon the large logarithms we wish to resum is encoded in the radiator:
\beq
R\left(\bar{b}M\right) = L g^{(1)}(\as L) + g^{(2)}\left(\as L\right) + \frac{\as}{\pi} g^{(3)}\left(\as L\right)\,,
\eeq
where $L=\ln(\bar{b}^2M^2)$ and $\as = \as(M)$.
The functions $g^{(i)}$ are the same as in~\cite{FlorenceQT} but in our case the radiator does not contain any term involving the DGLAP anomalous dimensions or the coefficient functions. These $N$-dependent contributions have been used to evolve the parton distribution functions or have been taken care of by evaluating the running coupling in front of the coefficient functions at the appropriate scale $1/b$. The coefficient $A^{(3)}$ appearing in the function $g_3$ has been recently determined~\cite{BecherNeubertA3}. However, we currently include in $g_3$ only the terms which are relevant at $O\left(\as^2 \right)$, and we call our partial NNLL resummation NNLL$^*$.

\begin{figure}
\begin{center}
  \includegraphics[height=.25\textheight]{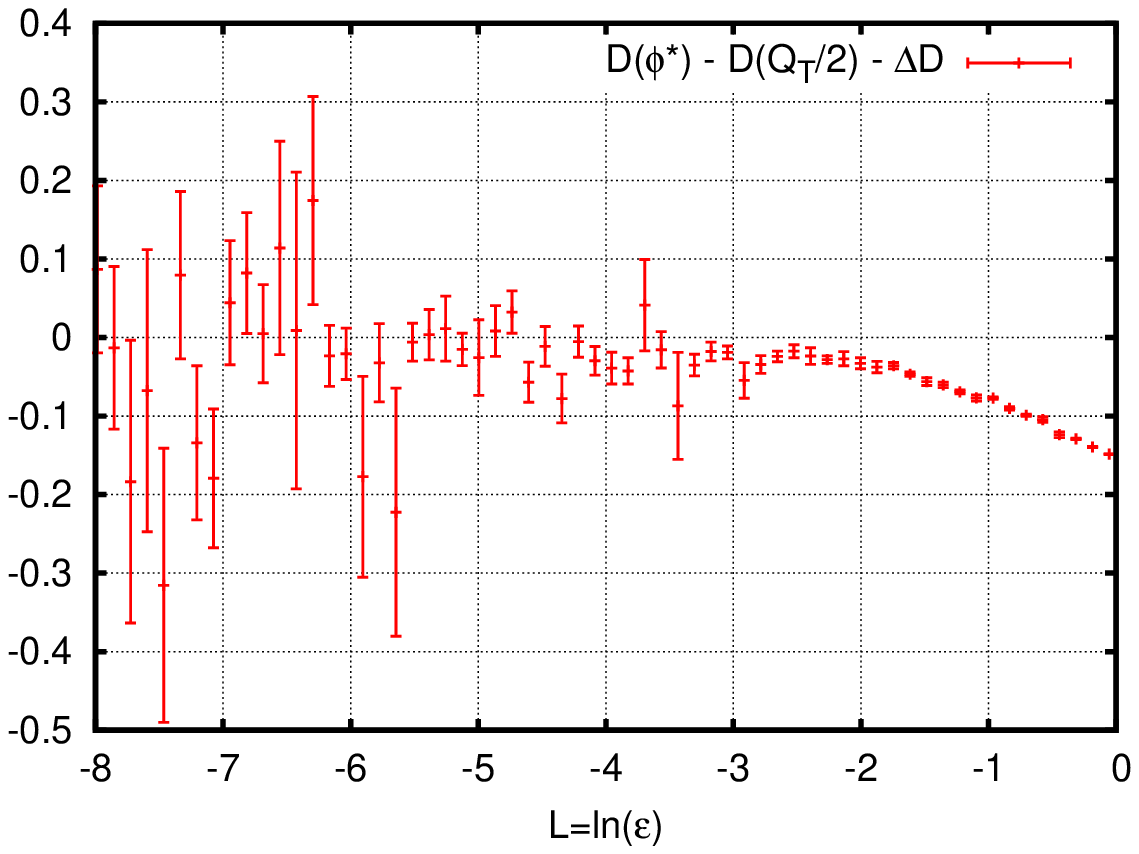}
    \includegraphics[height=.25\textheight]{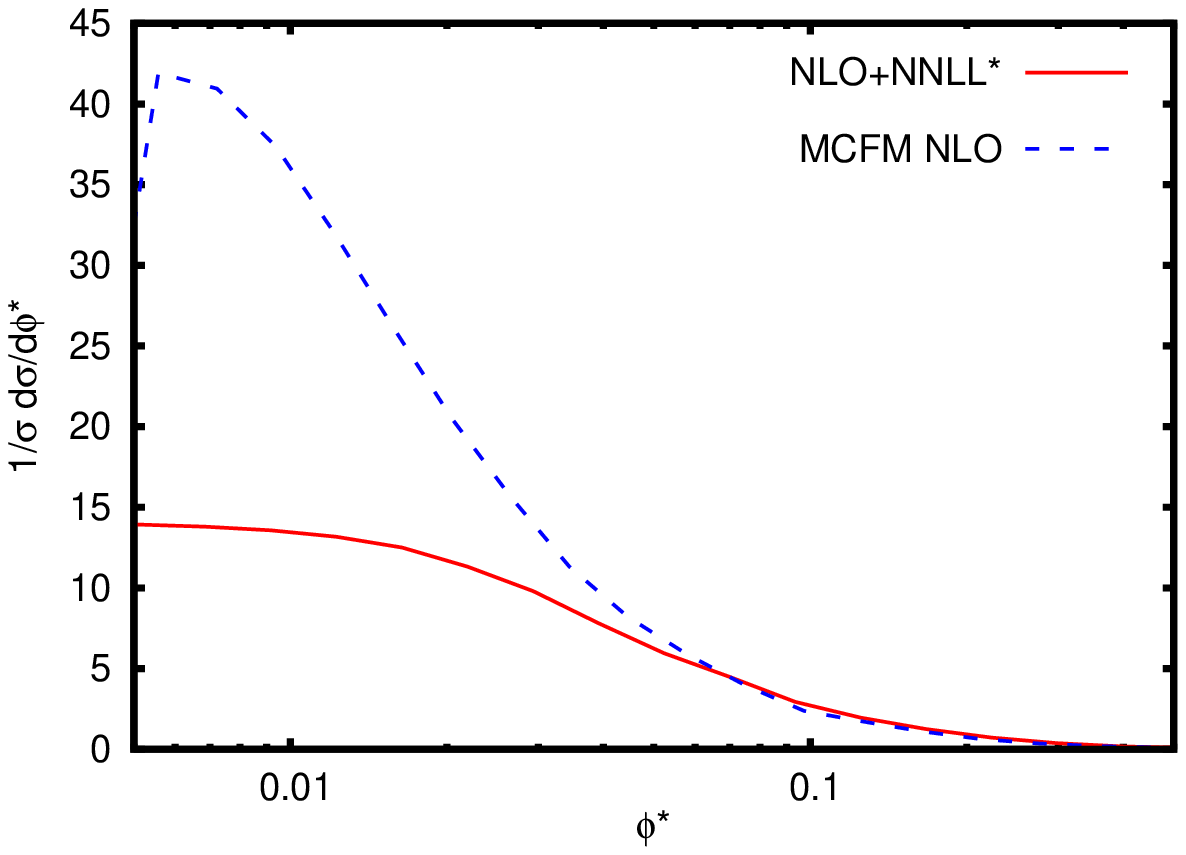}
\caption{On the left: the difference between the NLO differential distributions for $\phi^*$ and $Q_T/2$ from MCFM after removal of logarithmic terms from the resummation. On the right: the differential $\phi^*$ distribution computed at NLO with MCFM (dashed blue line) and our final result (solid red) obtained by matching the NNLL$^*$ resummation to the fixed order calculation at $O(\alpha_s^2)$, from MCFM. The distributions are normalised to NLO cross section.}  \label{fig:matching}
\end{center}
\end{figure}

Before presenting results for the matched distribution Eq.~(\ref{matched}) we must check that the expansion of our resummation to $O(\as^2)$ agrees with the NLO calculation. 
We remind the reader that the relation between the resummation for $Q_T$ and $a_T$ was worked out in~\cite{BDDaT} and we also have that $\phi^*\sim a_T/M$ as small $Q_T$.  In order to check our understanding of the relation between the different observables we compute
\begin{equation}
\Delta D(\epsilon) = \frac{1}{\sigma_0} \frac{{\rm d }}{{\rm d} \ln \epsilon} \left( {\sigma}\left(\phi^*\right)\big |_{\phi^*=\epsilon}-{\sigma} \left(Q_T/2\right) \big |_{Q_T/2=\epsilon} \right).
\end{equation}
Subtracting $\Delta D$ from the corresponding fixed-order differential distribution $D\left(\phi^*\right)-D\left(Q_T/2\right)$, computed with MCFM at NLO, we find a result that tends to zero, as shown in Fig.~\ref{fig:matching}, on the left.
Thus, we have complete control over the divergent pieces at
order $\alpha_s^2$ and we can adopt a particularly
simple matching formula like the one in Eq.~(\ref{matched}), where one adds the resummed differential distribution to the NLO result and subtracts the expansion of the resummation to order $\alpha_s^2$.

The result for the matched differential  $\phi^*$ distribution is plotted in Fig.~\ref{fig:matching} on the right, together with the pure fixed order $O\left(\alpha_s^2 \right)$ calculation obtained from MCFM.  The curves are obtained taking into account the D\O $ $ cuts for the muons, i.e. $70<M<110$~GeV, $p_T>15$~GeV and $\eta<2$, and integrating over the rapidity of the Z boson. 
We notice that the NLO calculation diverges in the region of small $\phi^*$, while the resummed and matched results tend to a constant, without forming any peak.
While low values of $\phi^*$ or $Q_T$ can be obtained via Sudakov suppression or kinematical cancellation,  in the present case of $\phi^*$ the kinematical cancellation starts to dominate prior to the formation of the Sudakov peak, in contrast to the $Q_T$ case.

\begin{figure}
\begin{center}
  \includegraphics[height=.25\textheight]{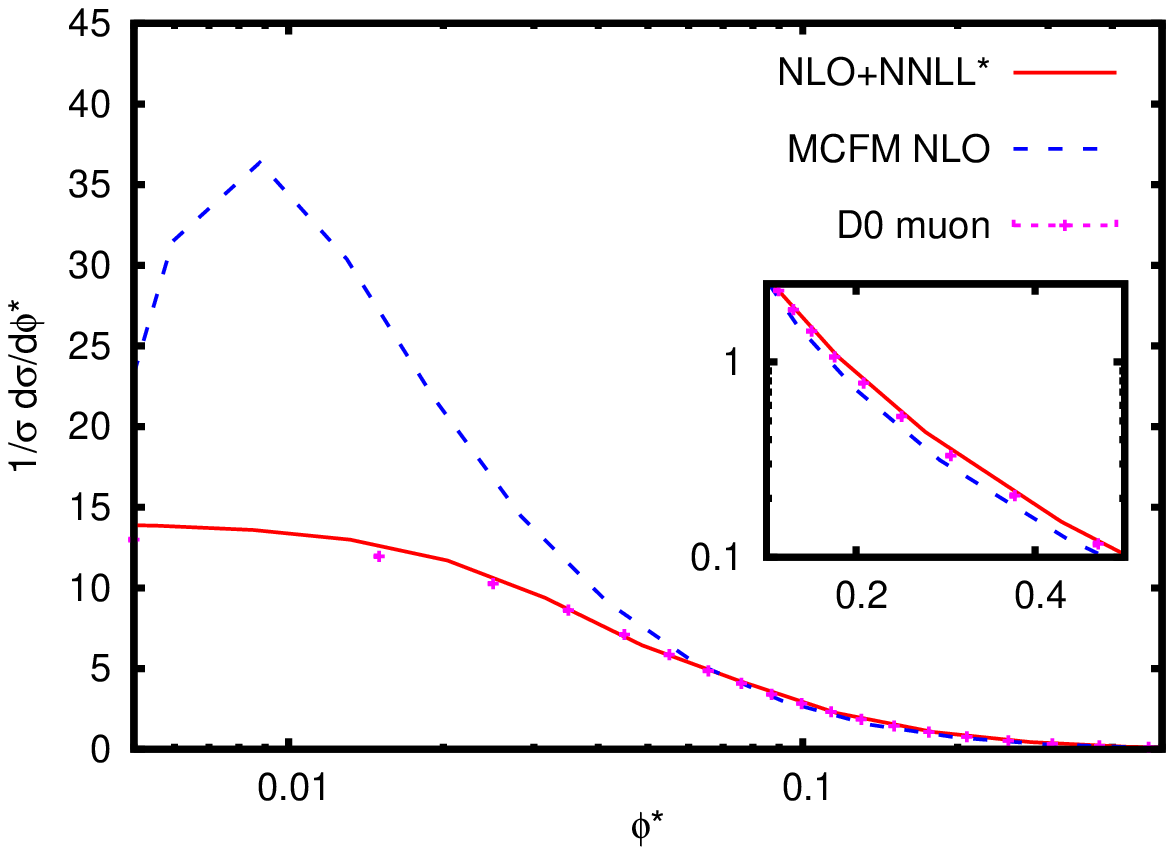}
    \includegraphics[height=.25\textheight]{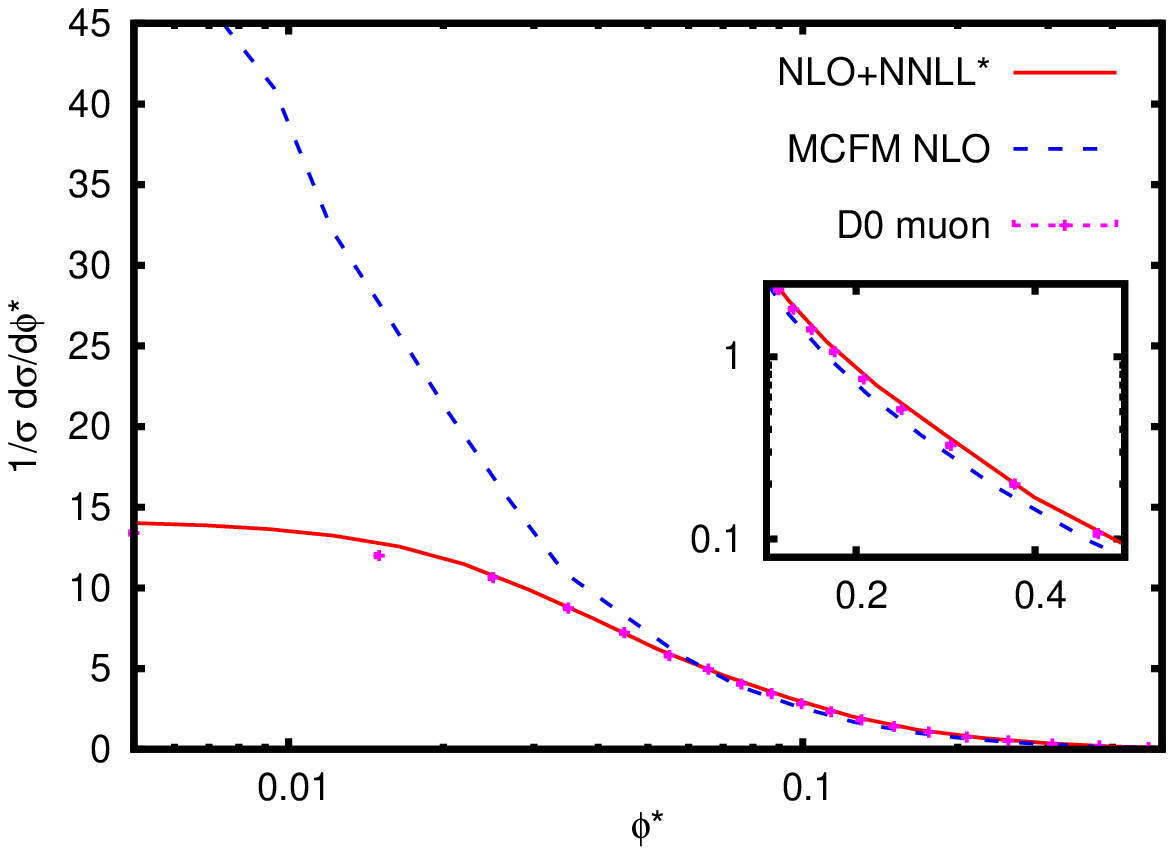}
\caption{Comparison of our resummed and matched theoretical prediction to the muon data collected by the D\O $ $ collaboration. The plot on the left corresponds to the central rapidity bin $|y|<1$, while the one on the right is for $1<|y|<2$. The curves have been obtained without non-perturbative effects.} \label{fig:D0muon}
\end{center}
\end{figure}

We are now in the position to perform a first comparison to the data obtained by the D\O $ $ collaboration~\cite{D0dphi}. For the moment we restrict ourselves to the muon case in two different rapidity bins: $|y|<1$ in Fig.~\ref{fig:D0muon}, on the left and $1<|y|<2$, on the right. We can clearly see that in both cases the matched result gives a good description of the data, with a maximum discrepancy of 7/8 \% in the last $\phi^*$ point. This is a very good result because our curves are obtained from a pure perturbative QCD calculation, with no other ingredients. The insets show that the matching has  beneficial effects also at large $\phi^*$. 

The plots in Fig.~\ref{fig:D0muon} are very encouraging but still not complete. First of all we have to include the complete expression for the function $g_3$. Moreover, before making any statement about the size on non-pertubative effects, we have to study the theoretical uncertainty. More precisely we can make an estimate of  missing higher orders in the fixed-order part by varying renormalisation and factorisation scales $\mu_R$ and $\mu_F$. We can also estimate missing higher logarithmic orders by varying the resummation scale $\mu_Q$. Moreover we need to investigate  different sets of parton distribution functions and different procedures to regularise the $b$-integral.  
Having done that,  we will be able to add an uncertainty band to our curves,  properly compare to the D\O $ $ data and pin down the non-perturbative contribution.

From a more theoretical viewpoint we note the resummation presented here is closely related to the one for $\Delta \phi$ between jets~\cite{BanDasDel08}, which therefore signifies an extension of $Q_T$ resummation to processes with colour in the final state.

\section{Conclusions}
We have reported on phenomenological work on accurate predictions for novel variables recently measured by the D\O $ $ collaboration to probe the low $Q_T$ region of the Z boson spectrum. Our result resums large logarithms at NNLL$^*$ accuracy and it is matched to a NLO calculation. The first comparisons to data are encouraging, although before drawing any conclusions about the size of non-perturbative effects, we must estimate the theoretical uncertainty.

\end{document}